\documentclass[aps,pra,twocolumn]{revtex4}

\usepackage{graphicx}
\usepackage{hyperref}

\begin{document}

\title{Experimental demonstration of frequency-degenerate bright EPR beams \\with a self-phase-locked optical parametric oscillator}

\author{Ga\"{e}lle Keller$^{\dag}$, Virginia D'Auria$^{\dag}$, Nicolas Treps$^{\dag}$, Thomas Coudreau$^{\ddag}$, Julien Laurat$^{\dag}$, Claude Fabre$^{\dag}$}

\address{$^{\dag}$ Laboratoire Kastler Brossel,
Universit\'{e} Pierre et Marie Curie, Ecole Normale Sup\'{e}rieure,
CNRS, Case 74, 4 place Jussieu, 75252 Paris Cedex 05, France}

\address{$^{\ddag}$Laboratoire Mat\'{e}riaux et Ph\'{e}nom\`{e}nes Quantiques, Universit\'{e} Denis Diderot, CNRS, \\10 rue A. Domon et L.
Duquet, 75205 Paris Cedex 13, France}

\begin{abstract}
We report the first experimental observation of bright EPR beams produced by a type-II optical
parametric oscillator operating above threshold at frequency
degeneracy. The degenerate operation is obtained by introducing a
birefringent plate inside the cavity resulting in phase locking.
After filtering the pump noise, which plays a critical role,
continuous-variable EPR correlations between the orthogonally polarized signal and idler beams are demonstrated.
\end{abstract}

 \maketitle

Beyond a fundamental significance, entanglement has proven to be a
necessary resource for quantum information protocols
\cite{zoller05}. In parallel to the photon-counting regime, an
active direction focuses on continuous variables of light
\cite{cerf}. In this context, different techniques have been used to
generate entangled beams. The more efficient ones rely on parametric
interaction in an optical cavity. Type-I parametric amplifiers
enable to generate squeezed states and mixing two of them on a
beam-splitter result in entangled beams
\cite{akira98,Bowen04,akirarecent}. Another way is to use type-II
amplifiers, which directly provide orthogonally-polarized entangled
beams \cite{Kimble92,Zhang00,Schori02,laurat}. These devices are now
widely used for generating deterministic entanglement and
entanglement of formation as high as 1.1 ebits has been obtained
\cite{laurat}.

Such non-linear systems can also be used above threshold. Although
strong quantum intensity correlations have been repeatedly obtained
with type-II optical parametric oscillators (OPO)
\cite{heidmann,gao,Laurat05b}, phase anticorrelations are difficult
to demonstrate experimentally. Above threshold, an OPO behaves as an
active oscillator, which chooses its working point providing the
lowest oscillation threshold. In particular, the signal and idler
beams are only accidentally at the same frequency. As a result,
homodyne detection cannot be implemented. Only very recently,
different groups working with type-II OPO managed to produce and
characterize bright entangled beams \cite{BRAZIL,Villar04,JING,SU}.
In these recent experiments, the beams were not at the same
frequency but complicated techniques of noise measurement have been
successfully used. We present here what is the first experimental
generation of bright frequency-degenerate EPR beams with an OPO,
characterized by homodyne detection.

The main device of the experimental setup is a type-II
self-phase-locked OPO, namely an OPO with an additional linear
coupling between the signal and idler fields. This device, initially
proposed by Mason and Wong \cite{mason} and later described by Fabre
et al. \cite{fabre}, has been extensively studied theoretically in
\cite{Longchambon04a} and its quantum properties have been detailed
in \cite{Longchambon04b}. The frequency-degenerate operation is
enabled by a classical, all-optical coupling between the signal
($A_{1}$) and the idler ($A_{2}$), induced by a birefringent plate
inside the OPO cavity, which forces the phase-locked operation of
the system within a finite range of parameter space. For very small
angles of the plate, it has been shown theoretically
\cite{Longchambon04b} that such a configuration preserves the strong
entanglement between the signal and idler modes predicted in an OPO
above threshold: they exhibit quantum correlations and
anti-correlations on orthogonal quadratures. Frequency degenerate
operation has been demonstrated in this system and enabled us the
use of homodyne detection. However, phase anticorrelations were
above the shot noise limit and prevented us from demonstrating
entanglement in this regime \cite{Laurat05b}. The main reason has
been identified as the pump excess noise. Its reduction and the
subsequent demonstration of entanglement and even of EPR
correlations is reported here.

\begin{figure*}[t!]
\begin{center}
\includegraphics[width=1.85\columnwidth]{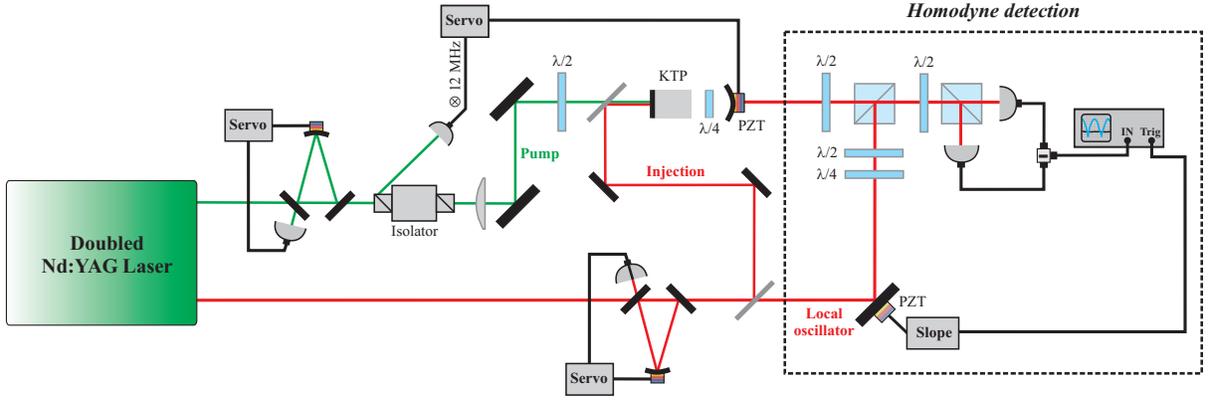}
\end{center}
\caption{The 532 nm output of a cw
frequency-doubled Nd:YAG laser is filtered
and used as the pump of a type-II OPO. The
OPO operates above threshold at frequency degeneracy thanks to a
quarter-wave plate inserted inside the cavity. The generated
two-mode state is characterized by homodyne detection.
}\label{Fig_Manip_article}\vspace{-0.4cm}
\end{figure*}

The experimental setup is sketched in Fig.
\ref{Fig_Manip_article}. The laser source is a continuous-wave
Nd:YAG laser (Innolight-\textit{Diabolo}) internally frequency
doubled. The output at 532 nm pumps a triply-resonant OPO, based on
a type-II 10 mm-long KTP crystal. The OPO is locked on the pump
resonance by the Pound-Drever-Hall technique (PDH). In order to
reduce the effect of the laser classical noise, the pump is filtered
by a 760 mm-long triangular cavity. This filtering cavity is made of
two flat mirrors with a reflectivity 97\% and a third mirror, with a
radius of curvature of 750 mm and a HR coating at 532 nm; the
corresponding linewidth is 3.5 MHz. The cavity is locked on the
maximum of transmission by PDH. In order to achieve better
mechanical stability and reduce losses, a semi-monolithic
configuration has been used for the OPO: the input cavity mirror is
directly coated on the crystal face. The input coupler is HR at 1064
nm and has a reflectivity of 95.5\% for the pump. The output mirror,
HR at 532 nm, has a reflectivity  at 1064 nm of 95\%, and a radius
of curvature of 38 mm. The crystal temperature is actively
controlled, with residual oscillation of the order of the mK. A
birefringent plate, $\lambda /4$ for the infrared and approximately
$\lambda $ for the pump, is inserted inside the cavity. It can be
finely rotated relative to the axis of the non-linear crystal by
steps of 0.01$^\circ$ thanks to a mount controlled by a
piezo-electric actuator (New Focus Model 8401 and picomotor): for
the results presented here, it is rotated by a very small angle
($\leq $0.02$^\circ$). The measured OPO threshold is 20 mW, with a
cavity bandwidth for the infrared around 50 MHz. To reach the
degenerate operation, adjustment of both the crystal temperature and
frequency of the pump is required. A small fraction of the laser
output at 1064 nm can be injected into the OPO and exploited for
finding the triple resonance. This beam is blocked while performing
the quadrature measurements on the OPO output.

To characterize its noise properties, the OPO output is sent to a
homodyne detection. A half-wave plate followed by a polarizing beam
splitter enables to choose the modes to be detected: the signal
$A_{1}$, the idler $A_{2}$, or the $\pm $45$^\circ$ rotated modes
$A_{\pm}$. The local oscillator (LO) is provided by the coherent
laser output at 1064 nm, filtered by a high finesse ($\sim$ 2500)
triangular cavity, locked on resonance by tilt-locking. The homodyne
detection is based on two balanced InGaAs photodiodes (Epitaxx
ETX300 without cap, quantum efficiency: 95\%). The visibility
reaches \ $\approx $ 0.98. A quarter-wave plate on the LO path
allows correcting the polarizing beam-splitters imperfections, while
the LO phase can be scanned thanks to a mirror mounted on a
piezoelectric actuator. The homodyne photocurrent is sent to a
spectrum analyzer (Agilent E4411B). The quadratures of the modes
$A_{\pm }$ and $A_{1,2}$ are measured for different values of the
noise analysis frequency, within the OPO bandwidth. All measurements
have been performed for a pump power approximately equal to 1.1
times the oscillation threshold.

Let us recall that entanglement between the orthogonally-polarized modes $A_{1}$ and
$A_{2}$ can be characterized in terms of the properties of the $\pm
$45$^\circ$ rotated modes defined by \cite{Laurat05a}:
\begin{equation}
A_{+}=\frac{A_{1}+A_{2}}{\sqrt{2}}\textrm{\quad and \quad
 }A_{-}=\frac{A_{1}-A_{2}}{\sqrt{2}}\textrm{.}
\end{equation}
Considering the noise spectrum of the sum or difference of signal
and idler fluctuations is strictly equivalent to considering the
noise spectrum of the rotated modes. For entangled modes $A_{1}$ and
$A_{2}$, the noise fluctuations of the modes $A_{\pm}$ are squeezed
below the standard quantum limit (normalized to $1$ in the
following), respectively on the phase quadrature
($Y_{+}$) and the orthogonal quadrature ($X_{-}$). For an OPO above threshold, and  a self-phase locked OPO in the limit of very small linear coupling, the noise variances can be written as
\begin{eqnarray}\label{eqnoise}
G_{X}&=&1-\frac{T}{T+\mu}\,\frac{1}{1+\Omega^2}
\\
G_{Y}&=&1-\frac{T}{T+\mu}\,\frac{1-2(V_0-1)(\sigma-1)}{\Omega^2+\sigma^2}
\end{eqnarray}
with $G_{X}=\left\langle \Delta X_{-}\right\rangle ^{2}$ and
$G_{Y}=\left\langle \Delta Y_{+}\right\rangle ^{2}$. $\sigma$
corresponds to the pumping parameter defined as the input pump
amplitude normalized to the threshold. $T$ stands for the
transmission of the output coupler, and $\mu$ for the extra-losses
in the cavity. $V_0$ is the phase noise of the pump normalized to
the standard quantum limit. In the ideal case, the signal and idler
beams generated by the OPO above threshold exhibit perfect intensity
correlations and phase anticorrelations, or, in other words, a
perfect noise suppression on the phase quadrature of the fields sum
$Y_{+}$ and simultaneously on the orthogonal quadrature of the
fields difference $X_{-}$. However, in the experiments, the presence
of classical phase noise on the pump beam can seriously affect the
phase anticorrelations as stated by eq. \ref{eqnoise}. Let us
underline that this limitation is not present below threshold where
the pump noise does not affect the output noise suppression.

\begin{figure}[t!]
\begin{center}
\includegraphics[width=.9\columnwidth]{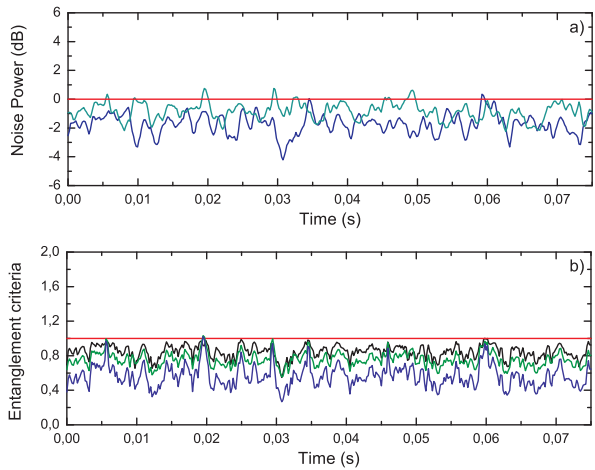}
\caption{(a) Normalized noise variances at 20 MHz of the $\pm
45^\circ$ modes and (b) entanglement criteria : Duan (green),
Mancini (violet) and EPR (black). The resolution bandwidth is set to
100 kHz and the video bandwidth to 300 Hz.}\label{Fig_Courbe20}
\end{center}
\end{figure}

Different criteria have been developed to prove entanglement \cite{Treps05}. The inseparability criterion developed by  Mancini \cite{Mancini02} or the one  by  Duan \textit{et al.} \cite{DUAN} and Simon \cite{SIMON} are  sufficient conditions for
$A_{1}$ and $A_{2}$ to be inseparable. They can be written respectively, in their simpler
experimentally testable form as:
\begin{equation}
G_{X}.G_{Y}<1 \textrm{\quad and \quad}
\frac{G_{X}+G_{Y}}{2}<1.
\end{equation}
The last quantity will be called separability in the following. The
third criterion is the EPR criterion, which characterizes a higher
degree of quantum correlations and is related to the situation
described in 1935 by Einstein, Podolsky and Rosen \cite{Einstein35}.
This criterion states that the beams are EPR-entangled if, by
performing a measurement on one of them, it is possible to retrieve
the values of two non-commuting observables on the other beam within
the quantum noise. The signal and idler beams exhibit EPR
correlations if they verify \cite{REID}:
\begin{equation}
V\left( X_{1}|X_{2}\right) V\left( Y_{1}|Y_{2}\right) <1.
\end{equation}
The quantity $V\left(A|B\right)$ corresponds to the conditional
variance on the observable $A$ obtained from a measurement of $B$.
In terms of the quantities $G_{X}$ and $G_{Y}$, the criterion reads
as:
\begin{equation}
\left( 2G_{Y}-\frac{G_{Y}^{2}}{\left\langle \Delta Y\right\rangle
^{2}}\right) \left( 2G_{X}-\frac{G_{X}^{2}}{\left\langle \Delta
X\right\rangle ^{2}}\right) <1
\end{equation}
with $\left\langle \Delta Y\right\rangle ^{2}$ and $\left\langle
\Delta X\right\rangle ^{2}$ the amplitude and phase variances for
the signal or the idler.

By scanning the local oscillator phase, one obtains the whole
quadrature noise properties of $A_{\pm }$. To
evaluate the entanglement criteria, one can hold the local
oscillator phase so that the measured quadrature would be the one
with the minimum noise. Figure \ref{Fig_Courbe20} (a) presents the
minimum quadrature noises of the rotated modes, normalized to the
shot noise, for an analyzing frequency of $20$ MHz, where it has been
measured that the pump excess noise can be neglected. Both
quadrature are squeezed : the noise suppression is $-0.8\pm 0.7$ dB
for $A_{+}$, and $-1.7\pm 0.8$ dB for $A_{-}$. The Duan
and the Mancini criteria can be directly evaluated from these
measurements (\textit{cf.} Figure \ref{Fig_Courbe20} (b) : one finds
a separability of $0.7\pm 0.1 < 1$, and $0.55\pm 0.10 < 1$ for the
Mancini criterion : the beams are indeed entangled. In order to
check for the more restrictive EPR criterion, the fields $A_{1}$ and
$A_{2}$ are also detected, as individual noises have to be taken into
account to determine conditional variances. As theoretically
predicted, the noise on these fields is phase-insensitive and they
are found to be approximately shot noise limited. These results
correspond to a value for the EPR criterion of $0.85\pm 0.10 <1$, thus
proving the  EPR type correlation between the signal and idler
beams. A summary is given on Table
1.

\begin{table}[b!]
\caption{Inseparability and EPR criteria, and experimental values
measured at 20 MHz.}\begin{center}
\footnotesize\begin{tabular}{|c|c|c|}\hline
 \textbf{Criterion} &\textbf{Expression} & \textbf{Measured}  \\\hline
  Mancini &  $ G_{X}.G_{Y}<1$ & $0.55\pm 0.10$ \\\hline
   Duan   &$(G_{X}+G_{Y})/2<1$ & $0.7\pm 0.1$  \\\hline
    EPR (Reid) & $  \left( 2G_{Y}-\frac{G_{Y}^{2}}{\left\langle \Delta Y\right\rangle
^{2}}\right) \left( 2G_{X}-\frac{G_{X}^{2}}{\left\langle \Delta
X\right\rangle ^{2}}\right) <1$ & $0.85\pm 0.10$  \\\hline
\end{tabular}
\end{center}
\end{table}

\begin{figure}[h!]
\begin{center}
\includegraphics[width=.9\columnwidth]{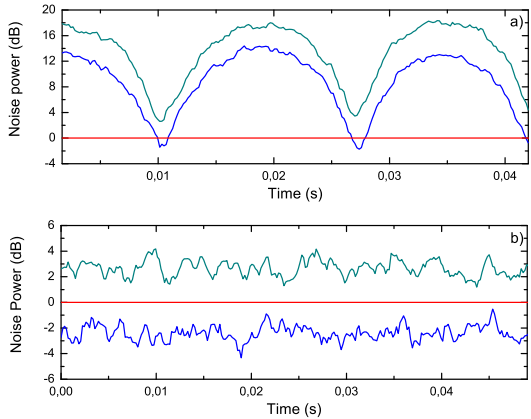}
\caption{Normalized noise variances at 3.5 MHz of the $\pm 45^\circ$
modes (a) while scanning the local oscillator phase and (b) while
locking it on the squeezed quadratures. The resolution bandwidth is
set to 100 kHz and the video bandwidth to 300
Hz.}\label{Fig_Courbe3v5}
\end{center}
\end{figure}

For further applications, it would be interesting to violate the EPR
criterion at lower frequencies. However, the residual phase
noise on the pump beam limits the noise suppression as stated by eq.
\ref{eqnoise}. To study this effect, we performed measurements at
different frequencies. Figure \ref{Fig_Courbe3v5} (a) presents
the quadrature noise properties of the modes $A_{\pm }$, normalized
to the shot noise as a function of the local oscillator phase, for
a frequency of $3.5$ MHz. For the mode difference $A_{-}$,
the curve minima are below the standard quantum limit; by locking
the LO phase on the squeezed quadrature, a noise reduction of
$-2.6\pm 0.3$ dB is measured (see Fig. \ref{Fig_Courbe3v5} (b). On
the mode $A_{+}$, the quadrature noise shows a phase dependence
similar to that observed on $A_{-}$, but it is always well above the
reference level. Separability and EPR criterion are both above one.
When increasing the frequency, the noise minimum reduces
and falls below the standard quantum noise from 6 MHz. There, a
squeezing of $-0.5\pm 0.5$ dB on $A_{+}$, and $-2.7\pm 0.7$ dB on
$A_{-}$ can be observed. These values correspond to an
inseparability of $0.7\pm0.1<1$, thus proving entanglement between
the signal and idler. The measured
 noise level is $+6.5\pm0.5$ dB both on $A_1$ and $A_2$,
which leads to conditional variances in the EPR criterion both
greater than 1: the pump excess noise is still too important at this
frequency. As a matter of fact, we did not detect EPR-correlations
below $20$ MHz. Let us note that the observed correlation at 20 MHz is in qualitative agreement with the results reported in ref. \cite{BRAZIL2} concerning the laser phase noise. Further experiments could reduce the low frequency
limit for entanglement detection, by improving the pump technical
noise suppression with a filtering cavity of higher finesse.

In conclusion, we have shown the ability of a compact device, a
type-II self-phase-locked OPO, to directly generate bright EPR beams
at the same frequency. The pump excess noise strongly affects the
entanglement and the pump thus needs to be filtered. We have
studied the effect of this noise and finally demonstrate EPR
correlations in a frequency region where it can be reduced. This
experiment explores a new regime about quantum properties of OPO
above threshold and provides a useful tool for quantum communication
protocols in the continuous variable regime.

This work was supported by the French ANR/PNANO contract IRCOQ.
Virginia D'Auria acknowledges financial support from La Ville de
Paris. We thank J.-A. Oliveira Huguenin for his work in the early stage of the experiment.

\end{document}